# Coherence of linear and surface data: Methodological proposal from the example of road data

Antonin Pavard[1], Patricia Bordin[2], and Anne Dony[1]

1 Université Paris-Est, Institut de Recherche en Constructibilité, Ecole Spéciale des Travaux Publics 28 avenue du Président Wilson, 94234 Cachan, France.
2 GeoSpective 16 B rue Charles Silvestri, 94300 Vincennes, France.

**Abstract:** The real world and its geographic objects are modeled and represented in different spatial databases. Each of these databases provides only a partial description (in space and time) of the geographic objects represented. Sometimes, producers and users of databases need to connect several of them for updates or comparisons. Much of the work in spatial database management focuses on matching these spatial databases and more particularly network databases, such as road networks. With regard to network data, one situation remains neglected, that of matching a linear database (with polylines objects) with a surface database (with polygons objects). In any case, users also need to connect these two types of spatial database. In this paper, a case study is made using French examples (Cachan, near Paris), as well as international case studies (Bordeaux in France, Victoria in Canada, and Copenhagen in Denmark), to propose an approach intended to make coherent network geographical objects in two different reference frames (linear and surface). This issue is addressed here through the example of road data. The surface data are then formalized in order to adapt them to linear data describing the same geographical objects. In the end, a polygon in the surface data corresponds to a single polyline in the linear data. This consistency should simplify the transfer of information from one reference frame (linear or surface) to the other. In other words, the methodology developed aims to make linear and surface geographic data interoperable.

# 1 Introduction

The real world is composed of *geographic objects* (roads, cities, etc.). In geographic information science, geographic information systems (GIS) enable the representation of geographic objects in their environment using two components [48].

First, a *spatial component* corresponds to the geometry (the geographic position and extent of the geographic object) and topology (the spatial relationships between objects) of an object. To represent the spatial component of a geographic object, we use what we call here *primitive objects* (a point, line, polygon, or raster cell). For example, the footprint of a city can be represented by a polygon.

Second, a descriptive component uses a set of *descriptive attributes* also called *alphanumeric attributes*. For example, a city can be described by its name or its population.

This information is structured and organized in tables. They are therefore databases to which spatial information is associated. These databases are called *spatial databases*.

The production of a spatial database is expensive and time-consuming. Yet, as early as the 1990s, researchers reported that the same geographic objects were being captured in several different spatial databases. They also predicted that this phenomenon would increase with time and the development of the Internet [53]. Today, this situation is real. Indeed, a wide range of spatial databases that describe the same geographic objects exists [56]. This diversity of databases relates to the variety of questions raised by specialists and data producers. Thus, different databases are created based on at least three aspects:

1) Different definitions of geographic objects to capture all their dimensions (environmental, societal, economic, etc.). For example, the geographic object "city" can be defined and represented according to institutional (political), morphological (urban), or functional (economic) criteria [47,11,42,7];

2) Different temporalities to study geographic objects at a given date or over time. For example, the temporal evolution of the geographic object "city" may be evaluated through several studies. These studies always underline the difficulty of having a single database to assess the evolution of all the cities in a territory [52,51,12,54,24];

3) Different spatial resolutions to study geographic objects at different scales. For example, the geographic object "city" maye studied on a small scale for questions of localization or city network analysis [57] or on a large scale for questions of spatial influence [25].

Ultimately, the multiplicity of spatial databases primarily derives from the difficulty of representing all dimensions of the same geographic object. The inclusion of all dimensions in a database makes the product more cumbersome and its exploitation more complex. However, both producers and users need to match spatial databases in order to link different dimensions of the same object. For producers, the challenge is to reduce the cost of updating the various databases; for users, the challenge is to benefit from diversified and rich data [33,28,49].

To address these challenges, various researchers have studied the feasibility of matching spatial databases and have highlighted *matching problems* [19,20,16,53]. Since then, this type of work has been applied to many different types of geographic object represented by spatial databases [37]. Network-type geographic objects, such as hydrographic networks and railroad networks, form the focal point of this paper, and in particular road networks.

Based on research undertaken for this paper, GIS applications realized on network-type geographic objects mainly concern the matching of spatial databases organized in the form of *graphs* [53]. This paper designates this type of matching as *graph-to-graph matching*. A graph is a combination of (i) *intersections*—also called *nodes*—represented by points and (ii) *sections*—also called *arcs*—represented by polylines that serve to connect the intersections.

This structuring favors a linear representation of the geographic objects represented. Thus, it is possible to benefit from the tools of graph theory to answer mobility questions [22].

Still, for network-type geographic data, GIS applications are less numerous or even non-existent when it comes to:

1) Matching two databases representing the footprints of these objects, i.e., using polygonal primitive objects. This paper designates this type of matching as *polygon-to-polygon matching*;

2) Matching between one database representing the footprints (surface data) and another representing the traffic axis (linear date organized in the graphs). This paper designates this type of matching as *graph-to-polygon matching*.

However, there are different technical needs or different themes for different GIS applications. These needs differ according to the geographic objects in question (hydrographic network, railroad network, road network, etc.). This paper takes the example of the road networks that were the subject of a previous study [43] and focuses on solutions for matching road databases.

First, it is worth noting that road geographic objects are most often represented using spatial databases structured as *graphs* (i.e., with point and polyline primitive objects). Indeed, the institutional and academic actors of the road most often work on road geographic objects seen as linear infrastructure. These actors are interested in the infrastructure in its spatial organization [34,2,8] or in its uses [50,25,23]. Their work is therefore based on graph-type spatial databases. This type of data is often associated with traffic or mobility information (volume, type, etc.).

Second, road geographic objects can also be represented, although less frequently, by spatial databases structured with polygonal primitive objects. Indeed, road actors also work on road geographic objects seen as infrastructure with a ground area to be developed. These actors are interested in road lanes and pavements, but also in all appendages to the road such as sidewalks and parking spaces. The development or maintenance of the road requires knowledge of the infrastructure's right-of-way in order to budget and plan the interventions in space [21]. Technical documents from the French Center for Studies and Expertise on Risks, the Environment, Mobility and Urban Planning (Cerema) show, for example, that the cost of road construction and maintenance depends on the surface and thickness to be covered [10]. The need for and existence of this type of representation are discussed in an earlier study [44]. Indeed, this type of representation of the road network geographic object is a recent development and mainly realized by local initiatives (GIS services of municipalities). This geographic data can be associated with information on road structures (materials or thicknesses).

Given the different ways of representing the road geographic objects, matching needs related to the three GIS applications presented above have been identified:

1) *Graph-to-graph matching* makes it possible either to compare databases [54,16,38,40,39] or to benefit from the latest updates proposed by the companies producing these data, such as Multinet for TomTom or NavteQ for Navstreet.

2) *Polygon-to-polygon matching* is hardly processed for road data today, if at all. This is because the polygonal representation of the road is still relatively new. Based on a previous study [44], a future need related to the existence of surface road databases representing the entire road right-of-way and other surface road databases representing the different road components (sidewalks, parking lots, etc.) can be hypothesized.

3) *Graph-to-polygon matching* is also hardly processed. Today, it arguably presents the greatest challenge in terms of GIS applications. The descriptive attributes of the traffic are associated with the road graph data, while the descriptive structure attributes can be associated with the road surface data. The structuring of a road is a *designing operation* [32]. This operation makes it possible to choose suitable materials (nature and thickness) and to associate them with the aim of designing or readjusting road infrastructure in order to make it durable over time. To make this choice, road specialists use several parameters including the traffic carried by the infrastructure. *Graph-to-polygon matching* therefore allows the transfer of descriptive attributes associated with polylines objects from graph databases to polygons objects from surface databases [45,43]. Beyond these public works, other specific data could be added; for example, research on heat islands has also highlighted the need to retrieve semantic information associated with graph databases to transfer it to surface databases, or vice versa [41,29]. Ultimately, it is the ability to conduct multiscale studies of both uses and infrastructure that is at stake in the interaction of these two types of road database.

This paper proposes a method to match two types of spatial database representing geographic objects organized in a network. This method is presented through the example of road data.

A method to make surface and linear road data consistent has been developed. This approach is designed from the analysis of "same as" spatial relationships as described in the first part of this paper. The approach is based on the construction of a surface data model, which has been built using a reference case study in the Paris metropolitan area (Cachan) and subsequently tested on other case studies in France (Bordeaux) and abroad (Victoria in Canada and Copenhagen in Denmark). This trial made it possible to test the model on different road configurations and data from heterogeneous sources. In order to facilitate understanding, the construction of the model is presented generically through simple cases. This paper ends with a discussion of the results, highlighting the successes and limitations of the method designed and implemented in order to propose possible areas of improvement.

## 2  Background

### 2.1  Relationships between spatial data

Mathematics is essential to explain the relationships between two sets: the *starting* set and the *ending* set. Each has several *elements*. Relationships are *applications* when all elements of the starting set relate to elements of the ending set. Figure 1 describes the three most common types of association in mathematics [35,1,6]:

1) The application is *one to one (bijective)* if each element of the ending set corresponds to one and only one element of the starting set. This relationship is written as *1-1*.

2) The application is *surjective* if each element of the ending set corresponds to at least one element of the starting set. This relationship is written as *1-n* or *n-1*.

3) The application is *multivalued* if each element of one set corresponds to several elements of the other set. This relation is written as *n-p*.

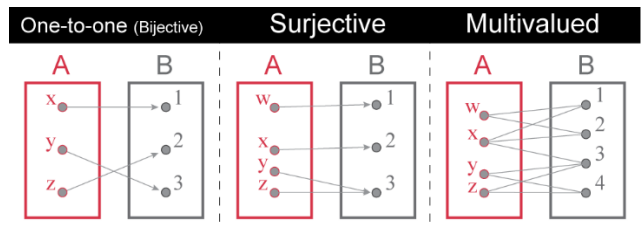

Figure 1: Three different applications

In computer science, and more particularly in *database management*, each set corresponds to a *database* composed of objects. For a spatial database that describes geographic objects, each computer object is represented by a *primitive object* (point, polyline, or polygon). The relationships between primitive objects can be described by their spatial relations, an area of research in computer science widely studied using different approaches [14,13], whether for spatial databases [26], geographic information systems [20,19], or image databases [3,4,30].

Clementini and Laurini [14] define several types of spatial relationship in spatial databases and GIS. This paper focuses on the relationships between objects that describe the same geographic object: the "same as" relationship. Based on this definition, the possibilities of matching two primitive objects from two different spatial databases is of interest.

## 2.2 Matching two primitive objects from two spatial databases

In accordance with the definition of spatial relationships in this paper, the applications described by mathematics are transposed to GIS (Figure 2). This makes it possible to present the most typical spatial relationships between primitive objects in two spatial databases carrying different semantic and attribute information:

1) *Type 1-1* allows an easy association from a spatial point of view. The spatial relationship is strong since a single primitive object in the *starting database* corresponds to a single object in the *arrival database*. Thus, the attributes describing the geographic objects and associated with the primitive objects can be transferred from one database to the other without difficulty. Precautions must be taken for the thematic validity of this transfer.

2) *Type 1-n* or *n-1* makes the spatial association more difficult. The spatial relationship is neither strong, nor weak, but medium since one primitive object from one database corresponds to several primitive objects in the other database. It is therefore necessary to build numerical transition rules to transfer the information describing the geographic objects and associated with the primitive objects of the databases.

3) *Type n-p* leads to the most difficult spatial association. The spatial relationship is weak since several primitive objects in one database partially correspond to several primitive objects in the other database. Thus, the transfer of information associated with primitive objects is even more complicated.

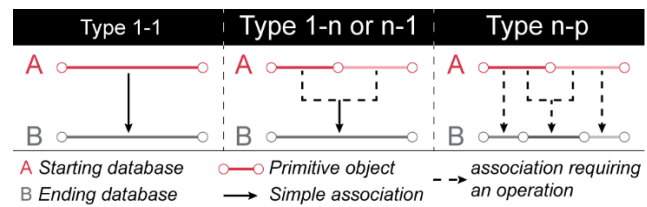

Figure 2: Translated applications for spatial databases

## 2.3 Matching polyline and polygonal primitive objects: Road databases case

This paper studies the spatial relationships between databases representing the geographic object "roads". Note that the road is located on the ground and is composed of *current sections*—also called road sections. These road sections connect the ends of dead ends and *road junctions* [9].

In the case of road data, the issue of association between polyline and polygonal primitive objects is particularly important at the level of road junctions. For the demonstrations in this paper, road interchanges and traffic circles have been excluded. From this, the four most common types of road junction in urban environments have been identified (Figure 3): cross, star, T, and Y [46,36].

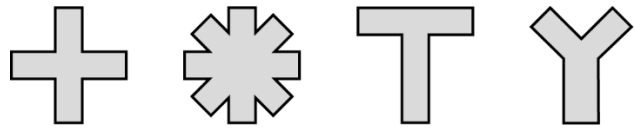

Figure 3: The four typical cases of road junctions

Indeed, T, Y, and star junctions are variations of cross road junctions. Thus, cross road junctions have been used to present the relationship between linear and surface road data.

Linear road databases organized as graphs are all structured in the same way. Polylines primitive objects that lead to the same road junction meet at the center point of this junction. In other words, the road junction is represented only by the junction of polylines primitive objects. By contrast, surface road databases are structured differently, with the biggest difference appearing in the representation of road junctions. In some cases, the road junction is represented distinctly by a primitive object, while in others it is aggregated to two current sections or to all current sections. These differences have an impact on the relationship between the two types of database:

1) When a road junction is modeled distinctly by an individual primitive object, the spatial relationship is almost always a type n-p relationship. The association is therefore difficult (Figure 4a).

2) When the road junction is aggregated to two current sections, the spatial relationship is a type n-1 or n-p relationship. The association is therefore medium (Figure 4b).

3) When the road network is represented by a single polygon, the spatial relationship is a type 1-n relationship. Several polylines are associated with one polygon. The association is also medium (Figure 4c).

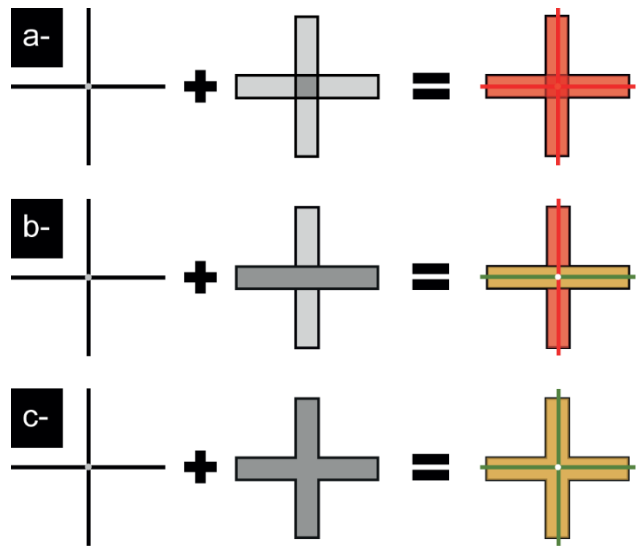

Figure 4: Spatial relationship between polylines objects and polygons objects

In all three cases, the association is limited. Some form of data processing is necessary and results in a loss of information (e.g., the choice of a percentage distribution). This paper proposes an alternative surface modeling method that takes into account the typical organization of linear road data. This modeling method is based on a division of road junctions according to the road sections leading to them, assigning part of the junction to each section. As a result, a 1-1 spatial relationship is obtained (Figure 5).

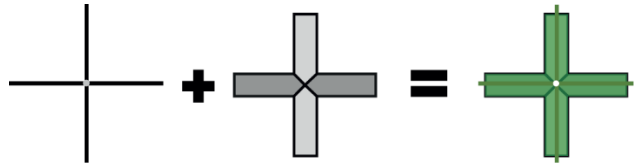

Figure 5: Proposed modelling

## 3   Materials and methods

### 3.1   Reference data

The objective of this paper is to develop a method to model polygons objects consistent with polylines objects representing the same geographic object. For this purpose, reference data in the form of a nominal case study is essential. A nominal case study is an observation of reality through filters defined by a specification at a given time. In other words, it is a picture of reality [15].

In the design of a modeling method, the nominal case study is used to verify the conformity of the results with data considered accurate, i.e., with reference data. These data are assumed accurate enough to measure and represent the objects of interest. Here, "reference datum" refers to data that is manually constructed in such a way as to respect the construction of the expected polygons objects (Figure 5).

This work was performed in Cachan, a territory well known to the authors of this paper because of its proximity to their research laboratory. Cachan is a French commune that developed as part of the expansion of Paris during the industrial revolution of the 19th century. Cachan is therefore a commune in the Paris metropolitan area located in the Val-de-Marne department, in the south of the "small ring" of the communes of Paris. Cachan is a small town with 30,200 inhabitants for 2.74 sq.km, that is to say, a density of approximately 11,000 inhabs./sq.km[1]. The road network represents 0.9 sq.km for 78 km linear. The history, location, and size of Cachan mean that it has a wide variety of urban spatial configurations, particularly in its road network (Figure 6a). The choice of Cachan to serve as the reference datum makes it possible to combine good knowledge of the case study with a variety of important road configurations, thereby ensuring the design of a modeling method that takes into account a maximum of situations.

This reference datum was compiled as part of previous work on the co-management of roads and technical networks, such as sewage systems or storm networks [43,44].

## 3.2   The case studies

In addition to the reference datum, three case studies were selected to test the modeling method from Cachan. These sites were chosen based on three criteria:

1) To have road data describing various configurations such as different types of junction to test specific modeling cases;

2) To use various data sources built by different organizations and freely available in order to test model reproducibility with different data;

3) To present applications in different national territories to verify the two preceding criteria.

The three case studies selected were Victoria (Canada), Bordeaux (France), and Copenhagen (Denmark).

The municipality of Victoria is the capital of the province of British Columbia in Canada. Victoria has an *orthogonal grid road system* typical of North American cities developed since the mid-19th century [5,27]. With an orthogonal grid, the forms of road junctions are simple (T or cross). Victoria is of a medium size with 85,792 inhabitants for 19.68 sq.km or approximately 4,359 inhabs./sq.km. The road network represents 4.5 sq.km for 268 km linear. Because of its size, it has significant road infrastructure, i.e., wide roads, sometimes connected by complex road junctions (Figure 6b).

The commune of Bordeaux is the capital of the Nouvelle Aquitaine region in France. Its history is marked by ancient Roman culture and the French Middle Ages as well as by the post-industrial and contemporary development of French territories. As such, it has a *winding road* in its historical center and a more varied but rectilinear road in its periphery [17]. Bordeaux is a slightly above average commune with 249,712 inhabitants for 49.36 sq.km or approximately 5,059 inhabs./sq.km. The road network represents 8 sq.km for 720 km linear. Because of its size, it has more road infrastructure than Victoria (Figure 6c).

[1] The information on populations for the case studies is provided by national statistical institutes in different countries. This information is given for the year 2016 for France and Canada and 2019 for Denmark.

The Municipality of Copenhagen is the capital of Denmark. Copenhagen is steeped in history, during which the Scandinavian peoples have left their mark. Like old French cities, the center of Copenhagen is organized around a winding road. With the rise of industry, the city began to expand like Paris in the 18th and 19th centuries. From the middle of the 20th century, the authorities, in consultation with architects and urban planners, chose to slow down urban expansion by implementing a "finger plan" [31]. This organization allowed for localized expansion along the railway lines while preserving green spaces. Today, Copenhagen is a large city with 623,404 inhabitants for 88.25 sq.km or approximately 7,064 inhabs./sq.km. The road network represents 16.7 sq.km for 1,316 km linear. The history and size of Copenhagen make it an interesting case study for this paper: residential areas are generally served by straight main roads, while the historical center sometimes has narrow and winding roads (Figure 6d).

## 3.3    Prerequisites: Input data

The development of the modeling method requires two sets of road data for the same territory:

1)   The first corresponds to a linear representation of the road network.

2)   The second corresponds to a surface representation of the same network.

The surface data must be organized in a single polygon representing the entire road network of the territory. In other words, the input is modeled road data without distinction between the road sections of the junctions (case illustrated in Figure 4c, p. 8). Victoria provided this type of data. In the other cases, an aggregation process must be performed:

1)   When road junctions and road sections are modeled separately as in Bordeaux (case illustrated by Figure 4a, p. 8);

2)   When the road junctions are associated with two road sections as in Copenhagen (case illustrated by Figure 4b, p. 8).

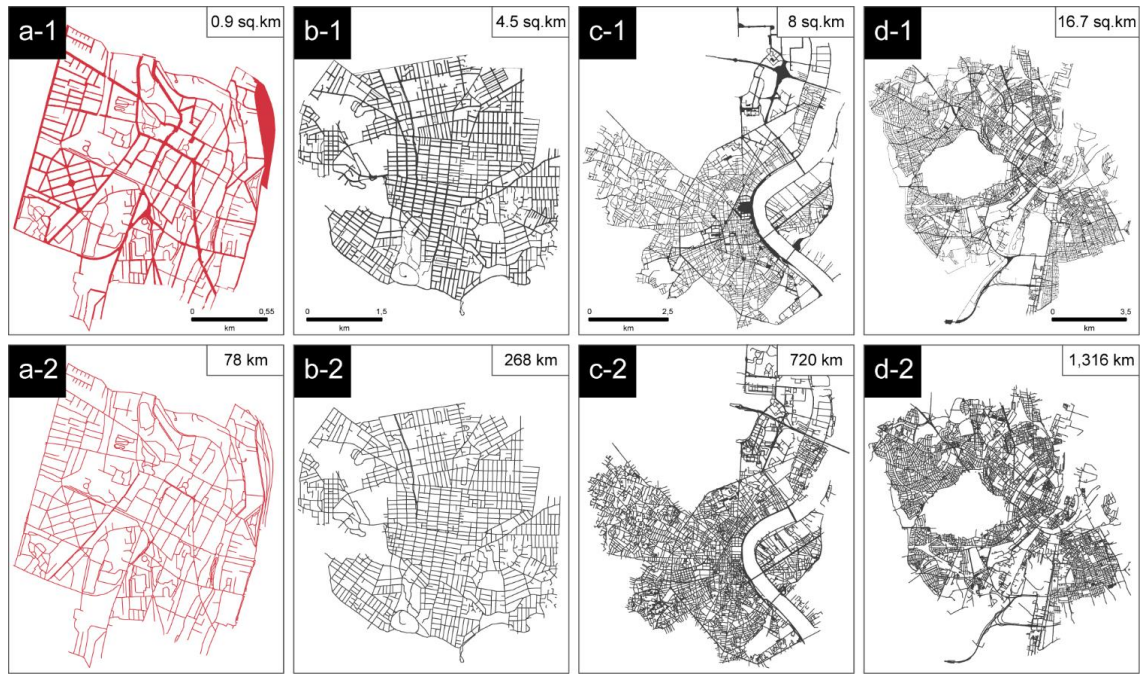

Figure 6: Road data. a) Cachan, b) Victoria, c) Bordeaux, d) Copenhagen, 1) surface data, 2) linear data. Sources: © IGN, BD TOPO® 2020, SIG Bordeaux Métropole 2019, CityOfVictoriaBC 2019, Københavns Kommune Bydata 2019

### 3.4 General principle of the method: Use of Voronoi tessellation

The proposed methodology is based on the division of polygons objects at the intersections of polylines objects. To do this, the plane was split into cells from a discrete set of points that have been called "source points."

The methodology in this paper is based on the concept of *tessellation*. Tessellation consists of covering the whole of a space with non-overlapping polygons geometric objects. In other words, the tessellation is a partition, which divides a given territory into regions (polygons). Each of these divisions contains the nearest *source point* at a Euclidean distance (Figure 7). The regions are formed by grouping all of the points closest to each source point.

As indicated by Dupuis [18], tessellation can take many forms. In this paper, *Voronoi tessellation* has been used. This is the most widely used tessellation in GIS, notably because of its genericity.

The methodology in this paper has been developed using an iterative process. At each stage, the results obtained were compared with the expected situation.

The expected situation corresponds to the allocation of a portion of the surface of a road junction to each road section that leads to it (Figure 5). These comparisons identified gaps. Hypotheses were then formulated to explain these shortcomings. Finally, improvements were proposed to correct them.

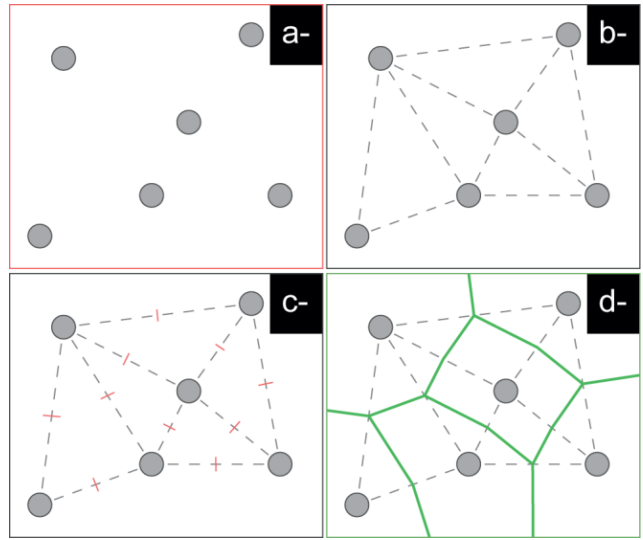

Figure 7: Tessellation process: From a set of points (a), each point is connected to its neighbors (b). The midpoints of each segment connecting the points are marked (c). Polygons are drawn from the midpoints to produce the tessellation (d).

### 3.5 Application of the method: Modeling of surface sections for optimal association

The objective of the method is to model the road surface sections, in order to allocate a part of the road junction to each road section (Figure 8).

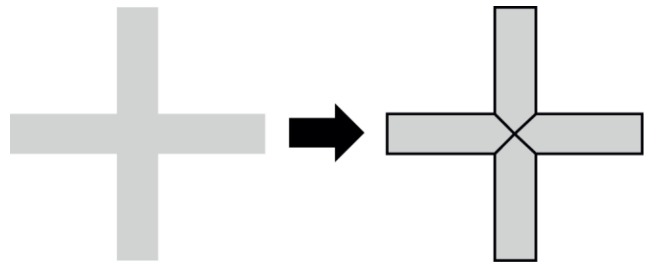

Figure 8: Modeling of surface sections of roads

To obtain this result, a *Voronoi diagram* has been used, built around points representing the linear intersection. Since a *Voronoi cell* is constructed for each source point, it is essential that each road junction has as many points as there are road sections leading to this junction. Constructing cells from a single point representing an intersection would result in encompassing the junction in a single Voronoi cell (Figure 9c1). Because of the way the cells in the Voronoi diagram are constructed, these points must be juxtaposed to the linear sections leading to the junction and be equidistant from the center of the junction. These points are called the buffer source points (Figure 9b). It is then possible to produce a homogeneous division of the road junction (Figure 9c2).

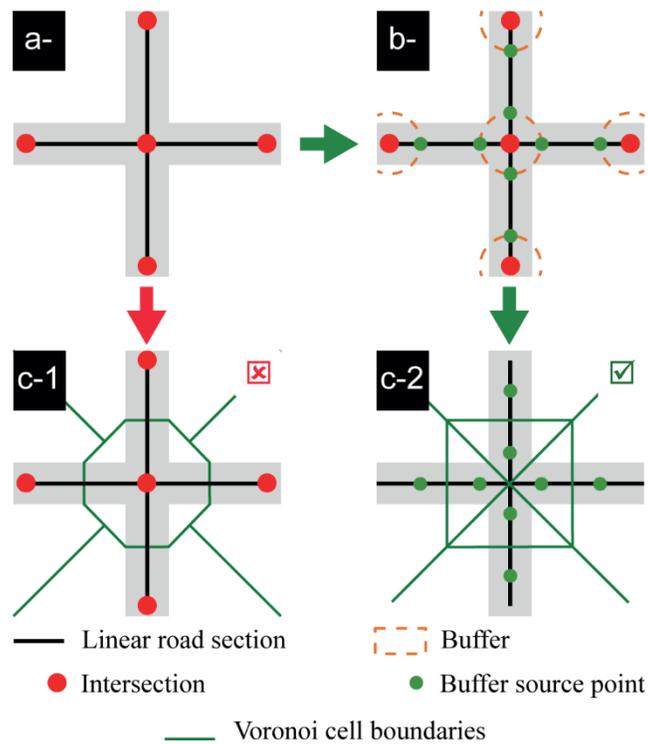

Figure 9: Construction of source points

In more concrete terms, a buffer zone is created around each *intersection point*. These buffer zones are then used to identify the buffer source points. The width of a roadway in an urban environment is generally between 5 and 7.5 meters [9]. In this paper, the minimum width of a roadway is used and therefore a radius of 5 meters is set for these buffer zones to best fit within the perimeters of junctions (Figure 10a). The buffer source points are constructed from the intersection of the polyline object and the buffer boundaries (Figure 10b), and the Voronoi diagram is calculated from the buffer source points (Figure 10c). A cut intersection is then obtained (Figure 8).

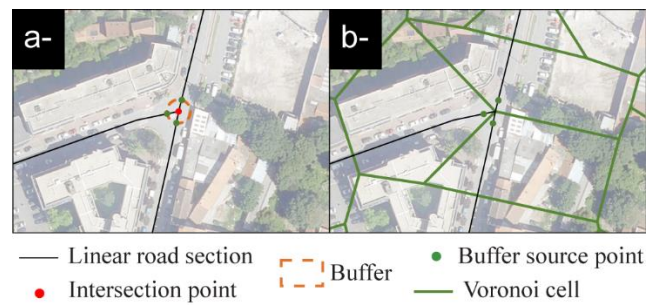

Figure 10: Voronoi diagram from buffer source points at a uniform distance to intersections. Sources: © IGN, BD TOPO®, ORTHO HR®, 2020

From the first results, several cases that cannot be simply treated by this process have been identified. These cases concern road configurations that arise commonly in dense urban environments: close junctions and close parallel roadways. Improvements to the method have therefore been proposed in order to make it equally efficient on these particular cases:

1) For close intersections, the radius of the buffer zone is determined in proportion to the length of the smallest road section leading to the junction and with a maximum radius of 5 meters.

2) For parallel roadways, *intermediate source points* are added along the length of the linear sections at intervals of 10 meters.

Finally, when the method is adjusted, road surface primitive objects (polygons) consistent with the linear road primitive objects (polylines) are produced (Figure 11).

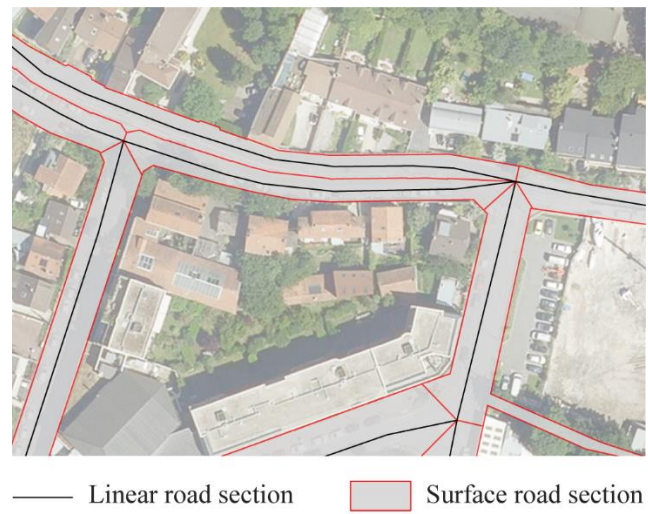

———— Linear road section     ☐ Surface road section

Figure 11: Final construction of the surface sections. Sources: © IGN, BD TOPO®, ORTHO HR®, 2020

## 4 Results and discussion

### 4.1 Comparison between modeled objects and reference objects

First, the results obtained by the method (modeled objects) were analyzed by comparing them with road surface sections produced by manual digitization (reference objects) on the reference case (Cachan). To do so, the surfaces of the reference road sections were compared with those produced automatically using a point cloud. Then a regression line was built (Figure 12).

The results of this analysis validate the approach taken. Indeed, a linear correlation with a slope close to 1 and a coefficient of determination of 87.25% can be observed. Moreover, the majority of the compared values are close to the regression line: 57% of the values have a surface deviation lower than 5%. This proportion increases to 67% for a deviation of 10% and to 73% for a deviation of 15%.

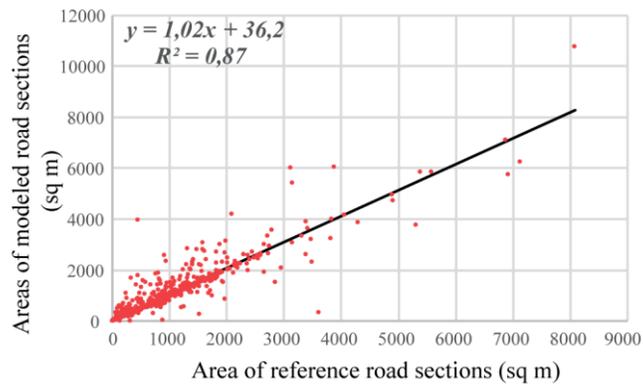

Figure 12 Comparison of modeled polygons objects with reference polygons objects

It has been assumed that the closer the surfaces of the modeled objects are to those of the reference objects, the more the method leads to the construction of consistent polygons objects. Overall, the modeled objects are similar to the reference objects. Beyond this observation, and to go further, the residuals, i.e., the objects for which the theoretical areas are the most distant from the observed areas, have been analyzed more precisely. These residuals fall into two categories:

1) First, despite regular corrections, splitting or unique identifier assignment can be observed in the manually generated dataset. These errors account for approximately 2% of all residuals that deviate most from the regression line.

2) Second, the road configuration makes it difficult to split the sections by area cleanly using a generic method. These splitting problems are presented in the next section. They result from three elements: the configuration of road junctions, the density of infrastructure, and the overlapping of roads when there is an engineered structure.

## 4.2   Presentation of modeled objects for all case studies

The developed method was then implemented on the three case studies (Victoria in Canada, Bordeaux in France, and Copenhagen in Denmark). These applications made it possible both to test the generic method across several territories and to show its adaptability to data of different types (Figure 13).

INPUT      OUTPUT

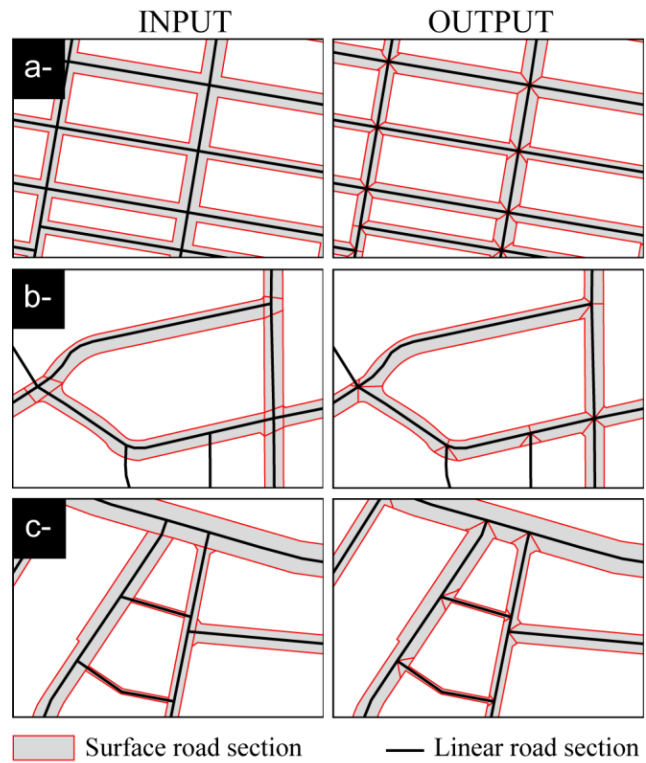

☐ Surface road section    —— Linear road section

Figure 13: Results of the implementation of the method on the three case studies: a) Victoria, b) Bordeaux, and c) Copenhagen. Sources: SIG Bordeaux Métropole 2019, CityOfVictoriaBC, 2019, Københavns Kommune Bydata, 2019

From these applications, it is worth noting the following:

1) The designed method is applicable as soon as one road linear geographic database and one road surface geographic database are available.

2) The time of realization depends on the volume of the input data[2]. For example, an area with less than 5 sq.km consisting of less than 100 km of linear road (e.g., Cachan in France, 2.7 sq.km per 100 km of road) currently requires approximately 30 minutes of processing time. For an area of approximately 100 sq.km and composed of a linear roadway of more than 400 km (e.g., Copenhagen in Denmark, 88.25 sq.km for 430 km of linear roadway), the process requires several hours.

3) The results for three road parameters (complex road junctions, density and proximity, and road overlap) are mostly correct but still open to criticism, and the method can therefore still be improved.

---

[2] The processing times indicated are given for a computer with the following configuration: Processor with 4 cores (1.9 to 2.1 Ghz) and 16.0 GB of RAM.

## 4.3    The results for the three road parameters

### 4.3.1    The complex road intersections

It can be observed that the type of junction and more specifically the angles and trajectories taken by the linear sections of the road have a direct impact on the shape of the resulting cut. It is worth noting that as the angles between the linear sections become more acute, the boundaries of the polygons objects become more irregular (Figure 14).

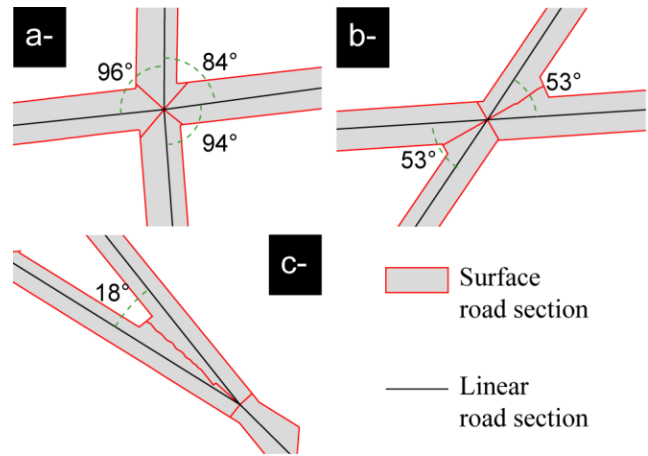

Figure 14: Examples of results for cross and T intersections. Sources: SIG – Bordeaux Métropole 2019; © IGN, BD TOPO® 2018

In addition to the angle, the curvature of the trajectories of linear primitive objects also changes the spatial relationship between the lines leading to the same junction. As soon as the trajectory of a linear section is curved, the angle formed between two linear sections is no longer fixed. The calibration of a regular cut then becomes more difficult and probably requires manual intervention (Figure 15). While this work can be done quickly for Copenhagen and Victoria where there are less than a dozen traffic circles, for example, it can become time-consuming for a region like Bordeaux where there are approximately 100 such junctions. An automatable process therefore still needs to be identified.

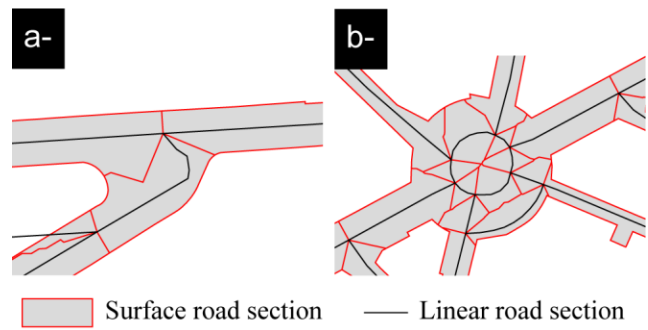

Figure 15: Modeling according to the shape or morphology of the road intersection.
Sources: SIG – Bordeaux Métropole 2019; © IGN, BD TOPO® 2020

### 4.3.2 The density and proximity of linear road sections

The treatment of current sections of road with a large right-of-way composed of parallel roadways remains problematic. The improvements proposed during the development of the method, such as the use of intermediate source points, have enhanced the results of these scenarios. However, a slight shift of the intermediate source points present on two parallel polylines objects can have an impact on the construction of the surface section boundaries.

A distance criterion has been chosen to determine whether or not to take into account the intermediate source points present on the neighboring linear sections. However, this criterion is perhaps not the most suitable to take into account these particular configurations correctly. For the same distance, two road sections may or may not belong to the same right-of-way (Figure 16a and 16b respectively).

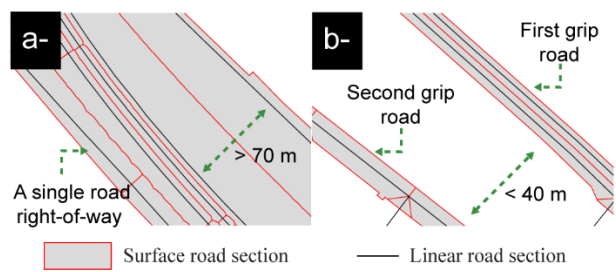

Figure 16: Linear density, special cases. Sources: SIG – Bordeaux Métropole 2019; © IGN, BD TOPO® 2020

To solve these cases, it would be interesting to replace the distance criterion by a contiguity criterion between road sections.

### 4.3.3 Road overlaps

The last point concerns overlapping road sections. The errors in these cases are not so much due to technical considerations as to the reasoning used when developing the method. Indeed, a planimetric road network was assumed, which is also most often the case in urban areas. However, whether or not a large proportion of the intersections are traffic circles, other junctions may represent types of interchanges, potentially in the form of the crossing of two *road sections* by superimposed pieces of infrastructure, or even engineering structures such as bridges or tunnels.

The method in this paper does not currently allow the creation of overlapping surface sections. However, two possible improvements can be suggested:

1)    For a simple engineering structure, such as an isolated bridge or tunnel, it is possible to automatically duplicate surface sections when two polylines objects overlap.

2)    For complex interchanges or complex junctions (Figure 17), manual processing is certainly still required. As for the traffic circle issue, the time for manual processing varies from one territory to another. For example, there are fewer than five complex junctions in Victoria whereas there are dozens in Bordeaux and Copenhagen.

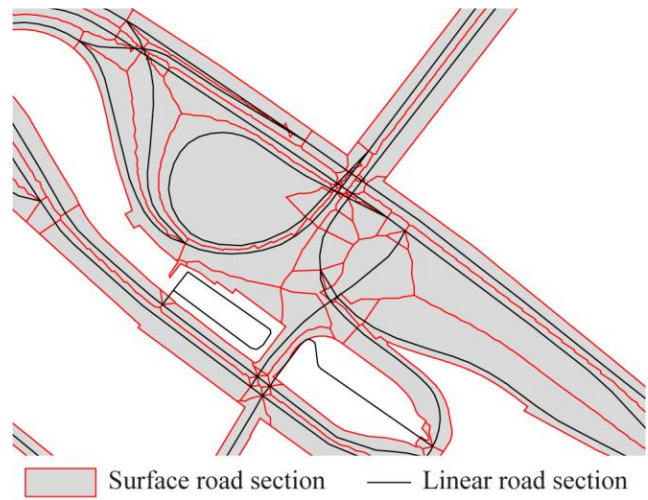

Figure 17: Case of motorway interchanges. Sources: SIG – Bordeaux Métropole 2019; © IGN, BD TOPO® 2020

## 5   Conclusion

This work aims to fulfill a need for interoperability between geographic databases that describe geographic objects organized in networks (road networks, hydrographic networks, etc.). To address this issue, this work has been based on the case of road networks.

In this paper, we started from different cases of polygonal primitive objects in order to show the limits of each one for a relationship with linear primitive objects. From there, we proposed a model that makes it possible to associate surface and linear data describing the same geographic object.

For road data, this modeling allows the transfer of data typically attached to linear representations, such as traffic data, to surface data as described, for example, by semantic data on the materials that constitute the road structures.

For example, this interaction of databases would make it possible to take into account the complexity of the road network. Indeed, road infrastructure evolves over time, whether through aging or through the impact of external phenomena, be they natural (e.g., climate impacts) or related to the infrastructure's purpose (e.g., road traffic impacts). It is therefore important to have all this information on a common surface reference.

Ultimately, this would make it possible to optimize the management of the road infrastructure.

In this paper, we have:

1) Proposed a model making linear and surface data interoperable;

2) Developed a method based on the road data of a reference case, the commune of Cachan, which has the advantage of having a wide range of road configurations;

3) Validated our approach and identified its limitations through statistical and visual analysis;

4) Demonstrated the reproducibility and genericity of this method by applying it to other data from heterogeneous sources and from different countries.

Thus, our method is applicable to the different types of road surface data identified, provided that linear road data organized in the form of a graph are also available.

The validation of the method and its reproducibility by applying it to very dense urban areas has highlighted particular difficulties. These relate to two main aspects:

1) Complex road junctions such as motorway interchanges:

   – Non-straightness of the road sections

   – Overlapping of road sections in the case of engineering structures (bridges, tunnels)

   – Road sections composed of several parallel carriageways

2) The linear representation of the infrastructure:

   – The accuracy of the modeling of the pavement axis

   – The more or less fine generalization of the lines.

For each of the specific difficulties identified, we propose potential ways of treating them. These special cases have been identified on the basis of the available data from cities in Western countries. We hypothesize that the morphology of cities in emerging countries, for which we do not have data, may present new specificities. We believe it would be interesting to extend this work to cities that may have different configurations.

# Acknowledgments

All authors contributed to the study conception and design. The original idea was proposed by Bordin P. Material preparation, data collection, and analysis were performed by Pavard A. The first draft of the manuscript was written by Pavard A., and all authors commented on previous versions of the manuscript. All authors read and approved the final manuscript. The research presented in this paper was financed by the foundation of the Ecole Spéciale des Travaux Publics (ESTP-Paris).